\documentclass[preprintnumbers,prd,twocolumn,showpacs,floatfix,preprintnumbers,superscriptaddress,nofootinbib]{revtex4-2}

\usepackage{graphicx}
\usepackage{epsfig}
\usepackage{bm}
\usepackage{amssymb}
\usepackage{float}
\usepackage{amsmath}
\usepackage{subfigure}
\usepackage{dcolumn}
\usepackage{booktabs}
\allowdisplaybreaks
\usepackage{multirow}
\usepackage[colorlinks]{hyperref}
\usepackage[usenames,dvipsnames]{color}
\hypersetup{
     breaklinks=true,
    pdfstartview={FitH},    
    colorlinks=true,       
    linkcolor=blue,          
    citecolor=red,        
    filecolor=magenta,      
    urlcolor=blue,           
    anchorcolor=green,      
    linktocpage=true
}
\usepackage{orcidlink}

\newcommand{\nn}{\nonumber}

\def\Om{\Omega}

\def\doi{http://doi.org}

\def\m{\mathrm{m}}

\allowdisplaybreaks

\begin{document}

\title{Probing the sensitivity of dark energy dynamics to equation of state parametrization flexibility}

\author{Md.~Wali~Hossain\orcidlink{0000-0001-6969-8716}}
\email{mhossain@jmi.ac.in}
\affiliation{Department of Physics, Jamia Millia Islamia, New Delhi, 110025, India}

\begin{abstract}
Allowing for greater low redshift flexibility through parametrizations such as CPL can lead to apparent deviations from $\Lambda$CDM. This motivates an investigation into whether such deviations reflect genuine dynamical dark energy or arise from parametrization choices. We investigate several dark energy parametrizations, including two phenomenological power law type extensions proposed here, the power law and modified power law forms, which allow greater low redshift flexibility (around $z \sim 1$) than CPL. Parametrizations permitting stronger evolution of the equation of state can exhibit enhanced phantom like behaviour at intermediate redshifts, but this feature is sensitive to the assumed functional form and its extrapolation. The statistical preference for such behaviour is modest, depends on the supernova sample, and is consistent across parametrizations, in agreement with recent analyses. The likelihood-ratio significance of this preference is $2.2$--$3.9\sigma$. Owing to limited data at $z > 2$, this is primarily driven by the equation of state in the range $1 \leq z \leq 2$, where parametrizations with enhanced flexibility, such as power law type forms, mildly improve the fit. The Akaike Information Criterion favours the DDE parametrizations over $\Lambda$CDM for all three supernova samples. The Bayesian Information Criterion mildly favours them for $+$Union3 and $+$DESY5 but disfavours them for $+$PantheonPlus.
\end{abstract}

\maketitle

\flushbottom

\section{Introduction}
\label{sec:intro}

The standard $\Lambda$ cold dark matter ($\Lambda$CDM) model, in which dark energy is described by a cosmological constant $\Lambda$, has been highly successful in explaining a wide range of cosmological observations, including the cosmic microwave background (CMB)~\citep{Planck:2018vyg}. However, increasingly precise data have revealed persistent tensions that challenge its internal consistency. Most notably, the Hubble constant inferred from Planck CMB measurements is in $\sim5\sigma$ tension with local determinations by the SH0ES collaboration~\citep{Riess:2021jrx,Planck:2018vyg}. Additionally, weak-lensing and large-scale structure measurements favour lower values of the clustering amplitude $S_8=\sigma_8\sqrt{\Omega_{\rm m0}/0.3}$ than predicted by the CMB~\citep{Perivolaropoulos:2021jda,Kilo-DegreeSurvey:2023gfr}. More recently, baryon acoustic oscillation (BAO) data from DESI, when combined with Type Ia supernova (SNe~Ia) and CMB observations, have been shown to favour dynamical dark energy (DDE) over $\Lambda$CDM at the $2.6\sigma$–$3.1\sigma$ level \citep{DESI:2024mwx,DESI:2025zgx,DESI:2025wyn,DESI:2025fii} within the Chevallier–Polarski–Linder (CPL) framework \cite{Chevallier:2000qy,Linder:2002et} (for related work, see~\cite{Giare:2024smz,Berghaus:2024kra,Wolf:2025jlc,Lee:2025pzo,Zhong:2025gyn,Yao:2025wlx,Qu:2024lpx,Wang:2024dka,Giare:2024gpk,Gialamas:2024lyw,Shlivko:2024llw,Ye:2024ywg,Bhattacharya:2024hep,Ramadan:2024kmn,Jiang:2024xnu,Payeur:2024dnq,Malekjani:2024bgi,Wolf:2023uno,Wolf:2024eph,Wolf:2024stt,Park:2024vrw,Park:2024pew,Dinda:2024kjf,Dinda:2024ktd,Jiang:2024viw,Colgain:2024xqj,Bhattacharya:2024kxp,Akthar:2024tua,Chan-GyungPark:2025cri,Ferrari:2025egk,Peng:2025nez,Colgain:2024mtg,Mukherjee:2024ryz,Mukherjee:2025myk,Mukherjee:2025ytj,RoyChoudhury:2024wri,Berbig:2024aee,Li:2024qso,Li:2024qus,Li:2025owk,Borghetto:2025jrk,Carloni:2024zpl,Liu:2025mub,Wolf:2025jed,Chudaykin:2024gol,Sohail:2024oki,Hossain:2025grx,Cheng:2025lod,Scherer:2025esj,Pedrotti:2025ccw,Escamilla:2023oce,Liu:2025myr,Lu:2025gki,Wolf:2025acj,Cline:2025sbt,Wang:2025dtk,VanRaamsdonk:2025wvj,Peng:2025tqt,Yang:2025oax,Jiang:2025hco,Ishak:2025cay,Shlivko:2025fgv,Silva:2025twg,Gialamas:2025pwv,Cortes:2025joz,Artola:2025zzb,Dinda:2025hiu,Park:2025fbl,Blanco:2025vva,Chaudhary:2025pcc,Chaudhary:2025vzy,Li:2025muv,Zhang:2025dwu,Gonzalez-Fuentes:2025lei,Gonzalez-Fuentes:2026rgu}). However, within the constant equation of state (EoS) cold dark matter ($w$CDM) framework, $\Lambda$CDM still lies well within the $1\sigma$ confidence region even with recent data \cite{DESI:2025zgx}. A noticeable deviation from $\Lambda$CDM appears only when more flexible low-redshift parametrizations, such as CPL, are considered, where $\Lambda$CDM lies more than $2\sigma$ away from the best-fit model. This behaviour suggests that present observations may prefer dark energy models with greater dynamical freedom than provided by simple constant-$w$ scenarios, thereby motivating a systematic exploration of dark energy parametrizations beyond $\Lambda$CDM \citep{Chevallier:2000qy,Linder:2002et,Barboza:2008rh,Dimakis:2016mip,Feng_2011,Ma:2011nc,Jassal:2004ej,Pan:2019brc,Colgain:2021pmf,Malekjani:2024bgi, Giare:2024gpk,Akthar:2024tua,Lee:2025pzo,Alam:2025epg}. Recent analyses by the DESI collaboration have extended this investigation using both parametric and non-parametric approaches \citep{DESI:2025fii}. These studies reveal consistent trends toward dynamical dark energy, including excursions of the EoS below $-1$ at intermediate redshifts, although the statistical significance remains moderate and the high-redshift behaviour is not yet tightly constrained.

In this context, DDE models, in which the EoS $w(a)$ evolves with cosmic time, provide a natural framework for addressing these observational tensions. Among the simplest and most widely studied examples is the CPL parametrization, $w(a)=w_0+w_{\rm a}(1-a)$, which offers a convenient and flexible description of dark energy dynamics at low redshifts, particularly for $z<1$. However, as discussed above, current observations tend to favour models in which the dark energy EoS possesses greater flexibility, motivating a systematic exploration of alternative parametrizations to examine whether a consistent pattern in this preference emerges that could guide us toward a more realistic description of dark energy. Nevertheless, the CPL form may be inadequate to capture the true dynamics of dark energy, especially for $z>1$, since it corresponds to a first-order Taylor expansion of $w(a)$, thereby further motivating the exploration of alternative parametrizations.

Recently, Ref.~\citep{Alam:2025epg} proposed a class of three-parameter dark energy EoS parametrizations. Due to the presence of an additional free parameter relative to CPL, these models achieve smaller minimum $\chi^2$ values for data combinations including all three SNe~Ia samples. Dynamically, they exhibit a more pronounced phantom phase than CPL for certain values of $(w_0,w_{\rm a})$, a behaviour that is also evident in the EoS evolution evaluated at the best-fit parameters. These findings suggest that certain parametrizations can exhibit enhanced phantom behaviour in their best-fit evolution, raising the question of whether this reflects a genuine physical preference or is influenced by the specific functional form adopted. If such behaviour were to be established with higher statistical significance, it could have several implications:
\begin{enumerate}
\item dark energy may be more phantom-like than previously inferred;
\item the CPL parametrization may become inadequate near $z\sim1$ or at higher redshifts;
\item this behaviour may guide the construction of parametrizations that better capture the true dynamics of dark energy, reducing arbitrariness in phenomenological modeling;
\item such models may exhibit increased statistical tension with the $\Lambda$CDM scenario.
\end{enumerate}

In this work, we investigate to what extent such phantom-like features are driven by the choice of parametrization and how robustly they are supported by current data. In addition to the CPL model, we consider two well studied two-parameter parametrizations, the Barboza–Alcaniz (BA) model~\citep{Barboza:2008rh} and the exponential (EXP) model~\citep{Dimakis:2016mip,Pan:2019brc}. To further extend our analysis, we introduce two additional, phenomenologically motivated, two-parameter forms, the power-law (PL) parametrization, $w(a)=w_0/a^\alpha$, and its generalization, the modified power-law (MPL) parametrization, $w(a)=2w_0/(1+a^\alpha)$. These parametrizations are introduced as phenomenological forms designed to explore how different functional extrapolations of $w(a)$ impact the inferred evolution. Despite having the same number of free parameters as CPL, they allow for distinct redshift dependence, particularly beyond the regime where low-redshift expansions are expected to be valid.

In the absence of strong theoretical guidance, the choice of parametrization introduces an element of arbitrariness. The goal of this work is therefore to assess how different parametrizations influence the inferred behaviour of the EoS and to what extent current data can robustly constrain such features.

\section{Models}
\label{sec:model}

As discussed in the introduction, we consider the well-studied CPL, BA, and EXP parametrizations, whose dark energy equations of state are given by
\begin{align}
    w(a) &= w_0 + w_{\rm a}(1-a) \,, 
    &&{\rm(CPL)} \,, \\
    w(a) &= w_0 + w_{\rm a}\,\frac{1-a}{a^2 + (1-a)^2} \,, 
    &&{\rm(BA)} \,, \\
    w(a) &= w_0 + w_{\rm a}\!\left(e^{1-a}-1\right) \,, 
    &&{\rm(EXP)} \,,
\end{align}
where $a$ is the scale factor.

The CPL parametrization provides a convenient first-order description of dark energy evolution around the present epoch, and is therefore most directly constrained at low redshifts ($z\lesssim1$). To allow greater freedom at higher redshifts, we introduce a simple extension of the $w$CDM model in which, instead of being a constant, the EoS evolves as a power law,
\begin{equation}
    w(a) = w_0\,a^{-\alpha} = w_0(1+z)^\alpha \, ,
    \label{eq:eos_para1}
\end{equation}
where $\alpha$ is a free parameter. For $\alpha>1$, this form diverges as $a\to 0$, indicating that it is intended as an effective low- to intermediate-redshift parametrization rather than a complete description of the early Universe. To regularize the high-redshift behaviour while retaining the enhanced flexibility at low and intermediate redshifts, we further introduce a modified power-law (MPL) form,
\begin{equation}
    w(a) = \frac{2w_0}{1+a^\alpha} \, ,
    \label{eq:eos_para2}
\end{equation}
which asymptotically approaches $w(a)\to2w_0$ as $a\to0$ and $\alpha>0$. This regularized behaviour allows one to explore the impact of different high-redshift extrapolations without conflicting with current observational constraints. While this fixed asymptotic value introduces a dynamical restriction, it has no practical effect on the present analysis, in the same way that the high-redshift divergence of the PL model does not, given the redshift range accessible to current data. Both parametrizations can be further generalized by introducing an additional parameter,
\begin{equation}
w(a)=\frac{(1+\beta)w_0}{\beta+a^\alpha} \, ,
\end{equation}
where $\beta$ controls the asymptotic behaviour of the EoS at high redshift. These parametrizations can be viewed as simple phenomenological extensions that probe how different functional forms for $w(a)$ influence the inferred dynamics, while maintaining the same number of free parameters as CPL. In this work, we restrict our analysis to the two representative cases $\beta=0$ and $\beta=1$, corresponding to the PL and MPL forms, respectively.

For $z\ll1$, both PL and MPL reduce to the CPL form with $w(a=1)=w_0$ and
\begin{align}
    w_{{\rm a},{\rm PL}} &= -\left.\frac{dw}{da}\right|_{a=1} = \alpha w_0 \,, 
    \label{eq:wa_PL}\\
    w_{{\rm a},{\rm MPL}} &= -\left.\frac{dw}{da}\right|_{a=1} = \frac{\alpha w_0}{2} \, .
    \label{eq:wa_MPL}
\end{align}
We stress that $w_{\rm a}$ in Eqs.~\eqref{eq:wa_PL}--\eqref{eq:wa_MPL} is not a free parameter for the PL and MPL models. It is fixed once $(w_0,\alpha)$ are chosen, and is quoted here only to allow direct comparison with the CPL, BA, and EXP parametrizations, for which $w_0$ and $w_{\rm a}$ are independently sampled.

\begin{figure*}[t]
\centering
\includegraphics[scale=0.55]{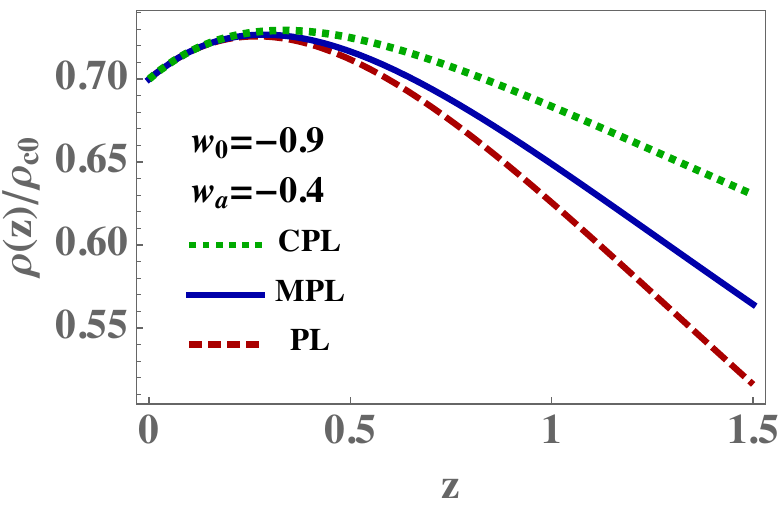}~~~~
\includegraphics[scale=0.55]{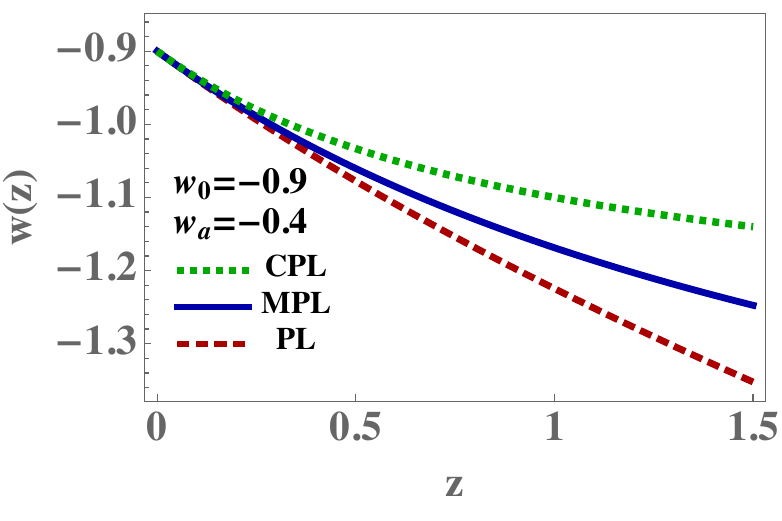}
\caption{Left: Dark energy density $\rho(z)/\rho_{\rm c0}$ is plotted against redshift $z$ for PL, MPL and CPL. Right: The corresponding EoS ($w(z)$) is plotted against $z$. Both panels use identical values of $w_0$ and $w_{\rm a}$ for all models.}
\label{fig:dyn}
\end{figure*}

The corresponding dark energy density evolution for the PL model is
\begin{equation}
    \rho_{\rm DE,PL}(z)=\rho_{\rm DE,0}(1+z)^3
    \exp\!\left[\frac{3w_0}{\alpha}\big((1+z)^\alpha-1\big)\right],
    \label{eq:rhoDE_para1}
\end{equation}
which reduces to the $w$CDM form in the limit $\alpha\to0$.  
For the MPL model, the dark energy density evolves as
\begin{equation}
    \rho_{\rm DE,MPL}(z)=\rho_{\rm DE,0}(1+z)^3
    \left(\frac{1+(1+z)^\alpha}{2}\right)^{\frac{6w_0}{\alpha}},
    \label{eq:rhoDE_mpl_final}
\end{equation}
recovering the $w$CDM scaling as $\alpha\to0$ and yielding $\rho_{\rm DE}\propto(1+z)^{3+6w_0}$ at very high redshifts, consistent with the asymptotic limit $w\to2w_0$.

Figure~\ref{fig:dyn} compares the evolution of $\rho(z)$, normalised with the present critical density $\rho_{\rm c0}$, and $w(z)$ for the PL, MPL, and CPL parametrizations using identical values of $w_0$ and $w_{\rm a}$. This comparison is intended to illustrate the qualitative differences in redshift evolution among the parametrizations under matched low-redshift conditions. While all models agree at very low redshifts ($z\ll 1$), the PL and MPL parametrizations rapidly deviate from CPL at higher redshifts, exhibiting a significantly enhanced phantom behaviour. This illustrates the restrictive nature of retaining only the first-order Taylor expansion in CPL and highlights the increased dynamical freedom of the power-law based models.

\begin{table*}[ht]
\centering
\caption{Cosmological parameter constraints (68\% CL) of model parameters of PL, MPL, CPL, BA, and EXP models using the +PantheonPlus, +Union3 and +DESY5 data combinations.}
\label{tab:params_constraints}
\resizebox{\textwidth}{!}{%
\begin{tabular}{lccccccc}
\toprule
\midrule
\textbf{Model}/Dataset & $\Omega_{m0}\times 10^2$ & $H_0$ & $\omega_b \times 10^3$ & $M$ & $r_{\rm d}/$Mpc & DE Parameters \\
\midrule
\textbf{$\Lambda$CDM} & & & & & & \\
+PantheonPlus & 30.28 $\pm$ 0.33 & 68.39 $\pm$ 0.26 & 22.51 $\pm$ 0.12 & $-19.411 \pm 0.008$ & 147.30 $\pm$ 0.17 & $w=-1$ (fixed) \\
+Union3 & 30.23 $\pm$ 0.33 & 68.43 $\pm$ 0.26 & 22.52 $\pm$ 0.12 & $\cdots$ & 147.31 $\pm$ 0.17 & $w=-1$ (fixed) \\
+DESY5 & 30.39 $\pm$ 0.33 & 68.30 $\pm$ 0.26 & 22.50 $\pm$ 0.12 & $\cdots$ & 147.27 $\pm$ 0.17 & $w=-1$ (fixed) \\
\midrule
\textbf{PL} & & & & & & \\
+PantheonPlus & 
31.14 $\pm$ 0.56 &
67.75 $\pm$ 0.59 &
22.43 $\pm$ 0.13 &
$-19.414 \pm 0.014$ &
147.05 $\pm$ 0.20 &
\begin{tabular}{l}
$\alpha=0.379 \pm 0.152$\\
$w_0=-0.874 \pm 0.046$ \\
$w_{\rm a}=-0.334 \pm 0.117$
\end{tabular} \\
+Union3 &
32.22 $\pm$ 0.79 &
66.72 $\pm$ 0.77 &
22.40 $\pm$ 0.13 &
$\cdots$ &
146.97 $\pm$ 0.20 &
\begin{tabular}{l}
$\alpha=0.653 \pm 0.211$\\
$w_0=-0.771 \pm 0.066$ \\
$w_{\rm a}=-0.502 \pm 0.123$
\end{tabular} \\
+DESY5 &
31.88 $\pm$ 0.55 &
67.03 $\pm$ 0.55 &
22.40 $\pm$ 0.13 &
$\cdots$ &
147.00 $\pm$ 0.20 &
\begin{tabular}{l}
$\alpha=0.547 \pm 0.161$\\
$w_0=-0.807 \pm 0.045$\\
$w_{\rm a}=-0.442 \pm 0.108$
\end{tabular} \\
\midrule
\textbf{MPL} & & & & & & \\
+PantheonPlus & 
31.10 $\pm$ 0.56 &
67.75 $\pm$ 0.59 &
22.43 $\pm$ 0.13 &
$-19.414 \pm 0.014$ &
147.09 $\pm$ 0.20 &
\begin{tabular}{l}
$\alpha = 0.853^{+0.650}_{-0.300}$\\
$w_0 = -0.873^{+0.066}_{-0.039}$ \\
$w_{\rm a} = -0.372^{+0.121}_{-0.237}$
\end{tabular} \\
+Union3 &
32.20 $\pm$ 0.71 &
66.64 $\pm$ 0.75 &
22.43 $\pm$ 0.13 &
$\cdots$ &
147.04 $\pm$ 0.19 &
\begin{tabular}{l}
$\alpha = 1.973^{+2.077}_{-0.644}$\\
$w_0 = -0.742^{+0.111}_{-0.062}$ \\
$w_{\rm a} = -0.732^{+0.201}_{-0.548}$
\end{tabular} \\
+DESY5 &
31.87 $\pm$ 0.56 &
66.97 $\pm$ 0.57 &
22.43 $\pm$ 0.13 &
$\cdots$ &
147.05 $\pm$ 0.19 &
\begin{tabular}{l}
$\alpha = 1.490^{+1.664}_{-0.368}$\\
$w_0 = -0.788^{+0.106}_{-0.036}$\\
$w_{\rm a} = -0.587^{+0.126}_{-0.492}$
\end{tabular} \\
\midrule
\textbf{CPL} & & & & & & \\
+PantheonPlus 
& 31.11 $\pm$ 0.56 
& 67.72 $\pm$ 0.59 
& 22.43 $\pm$ 0.13 
& $-19.414 \pm 0.015$ 
& 147.11 $\pm$ 0.20 
& \begin{tabular}{l}
$w_0=-0.856 \pm 0.056$ \\
$w_{\rm a}=-0.527 \pm 0.218$
\end{tabular} \\
+Union3 
& 32.38 $\pm$ 0.85 &
66.48 $\pm$ 0.83 &
22.40 $\pm$ 0.13 &
$\cdots$ &
147.03 $\pm$ 0.20 &
\begin{tabular}{l}
$w_0=-0.705 \pm 0.089$\\
$w_{\rm a}=-0.983 \pm 0.306$
\end{tabular} \\
+DESY5 
& $31.92\pm 0.56$ & $66.92\pm 0.56$ & $22.41 \pm 0.13$ & $\cdots$ & $147.06 \pm 0.20$ & 
\begin{tabular}{l} 
$w_0=-0.764 \pm 0.059$ \\ 
$w_{\rm a}=-0.802\pm 0.239$ 
\end{tabular} \\
\midrule
\textbf{BA} & & & & & & \\
+PantheonPlus 
& 31.09 $\pm$ 0.56
& 67.74 $\pm$ 0.59
& 22.44 $\pm$ 0.13
& $-19.414 \pm 0.014$
& 147.12 $\pm$ 0.20
& \begin{tabular}{l}
$w_0=-0.876 \pm 0.048$\\
$w_{\rm a}=-0.260 \pm 0.106$
\end{tabular} \\
+Union3
& $32.26^{+0.16}_{-0.94}$
& $66.59^{+0.94}_{-0.17}$
& 22.41 $\pm$ 0.13
& $\cdots$
& 147.05 $\pm$ 0.20
& \begin{tabular}{l}
$w_0=-0.753^{+0.060}_{-0.093}$\\
$w_{\rm a}=-0.460^{+0.182}_{-0.043}$
\end{tabular} \\
+DESY5 
& $31.86 \pm 0.55$ & $66.97 \pm 0.56$ & $22.42 \pm 0.13$ & $\cdots$ & $147.08 \pm 0.20$ & 
\begin{tabular}{l} 
$w_0=-0.800 \pm 0.049$ \\ 
$w_{\rm a}=-0.380 \pm 0.112$ 
\end{tabular} \\
\midrule
\textbf{EXP} & & & & & & \\
+PantheonPlus 
& 31.11 $\pm$ 0.56
& 67.74 $\pm$ 0.59
& 22.44 $\pm$ 0.13
& $-19.414 \pm 0.014$
& 147.09 $\pm$ 0.20
& \begin{tabular}{l}
$w_0=-0.868 \pm 0.051$\\
$w_{\rm a}=-0.395 \pm 0.166$
\end{tabular} \\
+Union3
& 32.29 $\pm$ 0.82
& 66.59 $\pm$ 0.81
& 22.41 $\pm$ 0.13
& $\cdots$
& 147.01 $\pm$ 0.20
& \begin{tabular}{l}
$w_0=-0.738 \pm 0.080$\\
$w_{\rm a}=-0.713 \pm 0.227$
\end{tabular} \\
+DESY5
& $31.89\pm 0.55$
& $66.97 \pm 0.56$
& $22.42\pm 0.13$
& $\cdots$
& $147.04\pm 0.20$
& \begin{tabular}{l}
$w_0=-0.787 \pm 0.054$\\
$w_{\rm a}=-0.589\pm 0.180$
\end{tabular} \\
\bottomrule
\end{tabular}}
\end{table*}

\begin{table*}[t]
\centering
\caption{Statistical model comparison between $\Lambda$CDM and DDE models (PL, MPL, CPL, BA, and EXP). $\sigma_{\rm LR}$ is the equivalent Gaussian significance of $\Delta\chi^2_{\rm min}$ for $k=2$ degrees of freedom.}
\label{tab:params_stats}
\resizebox{\textwidth}{!}{%
\begin{tabular}{lccc|ccccc|ccccc}
\toprule
\midrule
\multirow{2}{*}{Dataset} 
& \multicolumn{3}{c|}{$\Lambda$CDM} 
& \multicolumn{5}{c|}{PL}
& \multicolumn{5}{c}{MPL} \\
\cmidrule(lr){2-4} \cmidrule(lr){5-9} \cmidrule(lr){10-14}
& $\chi^2_{\rm min}$ & AIC & BIC
& $\chi^2_{\rm min}$ & $\Delta\chi^2$ & $\Delta$AIC & $\Delta$BIC & $\sigma_{\rm LR}$
& $\chi^2_{\rm min}$ & $\Delta\chi^2$ & $\Delta$AIC & $\Delta$BIC & $\sigma_{\rm LR}$ \\
\midrule
$+$PantheonPlus 
& 1436.06 & 1444.06 & 1465.66
& 1427.58 & -8.47 & -4.47 & 6.33 & 2.45
& 1427.98 & -8.08 & -4.08 & 6.72 & 2.37 \\
$+$Union3 
& 60.17 & 66.17 & 72.83
& 47.03 & -13.14 & -9.14 & -4.70 & 3.19
& 47.67 & -12.50 & -8.50 & -4.06 & 3.10 \\
$+$DESY5 
& 1688.15 & 1694.15 & 1710.76
& 1669.67 & -18.48 & -14.48 & -3.41 & 3.90
& 1669.94 & -18.21 & -14.21 & -3.14 & 3.86 \\
\bottomrule
\end{tabular}}
\resizebox{\textwidth}{!}{%
\begin{tabular}{lccccc|ccccc|ccccc}
\midrule
\multirow{2}{*}{Dataset} 
& \multicolumn{5}{c|}{CPL}
& \multicolumn{5}{c|}{BA}
& \multicolumn{5}{c}{EXP} \\
\cmidrule(lr){2-6} \cmidrule(lr){7-11} \cmidrule(lr){12-16}
& $\chi^2_{\rm min}$ & $\Delta\chi^2$ & $\Delta$AIC & $\Delta$BIC & $\sigma_{\rm LR}$
& $\chi^2_{\rm min}$ & $\Delta\chi^2$ & $\Delta$AIC & $\Delta$BIC & $\sigma_{\rm LR}$
& $\chi^2_{\rm min}$ & $\Delta\chi^2$ & $\Delta$AIC & $\Delta$BIC & $\sigma_{\rm LR}$ \\
\midrule
$+$PantheonPlus 
& 1428.86 & -7.20 & -3.20 & 7.61 & 2.21
& 1428.69 & -7.37 & -3.37 & 7.43 & 2.24
& 1428.11 & -7.94 & -3.94 & 6.86 & 2.35 \\
$+$Union3 
& 47.50 & -12.67 & -8.67 & -4.24 & 3.13
& 47.77 & -12.40 & -8.40 & -3.96 & 3.09
& 47.26 & -12.91 & -8.91 & -4.47 & 3.16 \\
$+$DESY5 
& 1670.22 & -17.93 & -13.93 & -2.85 & 3.83
& 1670.55 & -17.60 & -13.60 & -2.53 & 3.79
& 1669.88 & -18.27 & -14.27 & -3.19 & 3.87 \\
\bottomrule
\end{tabular}}
\end{table*}

\section{Data analysis}
\label{sec:data_analysis}

We constrain $\Lambda$CDM and five dark energy models, CPL, PL, MPL, BA, and EXP, using the latest cosmological observations described below.

\subsection{Observational Data}
\label{sec:data}

Our analysis combines multiple cosmological probes, including CMB, BAO, SNe~Ia and Hubble parameter data. The likelihoods for each dataset are given in Appendix~\ref{sec:Likelihoods}. 

For the CMB, we use distance priors reconstructed from the Planck 2018 TT, TE, EE+lowE measurements \citep{Planck:2018vyg,2019JCAP}. These compressed likelihoods capture the primary information from the CMB relevant for late-time cosmology while significantly reducing computational cost. Although such priors do not include the full information content of the Planck likelihood, they provide an accurate and widely used approximation for dark energy studies focused on background evolution \citep{2019JCAP}.  

The BAO dataset is taken from DESI DR2 \citep{DESI:2025zgx}, comprising over 14 million extragalactic objects, including emission line galaxies (ELGs), luminous red galaxies (LRGs), quasars (QSOs) \citep{DESI:2025qqy}, and Ly$\alpha$ forest tracers \citep{DESI:2025zpo}. These measurements provide $D_M/r_d$ and $D_H/r_d$ over $0.4 < z < 4.2$, along with low-redshift isotropic measurements of $D_V/r_d$ for $0.1 < z < 0.4$, thereby covering both isotropic and anisotropic BAO analyses.  

For type Ia supernovae, we use the PantheonPlus compilation \citep{Scolnic:2021amr,Brout:2022vxf}. We impose a cut $z>0.01$ on this redshift to mitigate the impact of local peculiar velocities and large-scale inhomogeneities. The Cepheid host (calibrator) SNe are also removed from the data vector and the covariance matrix. The resulting sample contains 1590 SNe Ia luminosity distance measurements, compared with the full PantheonPlus sample of 1701 measurements spanning $0.001 < z < 2.26$. We also include the Union3 compilation \citep{Rubin:2023ovl}, which provides 22 binned data points from 2087 cosmologically useful SNe~Ia, and the DESY5 sample \citep{DES:2024jxu}, consisting of 1635 SNe~Ia in the range $0.10 < z < 1.13$, supplemented by 194 low-redshift SNe~Ia ($0.025 < z < 0.10$), for a total of 1829 data points. We follow the latest DES-Dovekie recalibration of the DESY5 data \cite{DES:2025sig}.

We further include 31 measurements of the Hubble parameter $H(z)$ spanning $0.07 \leq z \leq 1.965$, obtained using the cosmic chronometer (CC) method~\citep{2018JCAP...04..051G}. 

We consider three data combinations corresponding to different SNe~Ia samples. The combination CMB$+$DESI DR2 $+$H(z)$+$PantheonPlus is denoted as $+$PantheonPlus. Similarly, CMB$+$DESI~DR2$+$H(z)$+$Union3 is referred to as $+$Union3, and CMB $+$DESI~DR2$+$H(z)$+$DESY5 as $+$DESY5.

\subsection{Priors and Model Parametrization}

For the $\Lambda$CDM model, we vary four cosmological parameters, $\{\Omega_{\rm m0},\, h,\, \omega_{\rm b},\, M\}$ for $+$PantheonPlus data set and three parameters $\{\Omega_{\rm m0},\, h,\, \omega_{\rm b}\}$ for $+$Union3 and $+$DESY5 data sets. Here, $\omega_{\rm b}=\Omega_{\rm b0}h^2$ is the physical baryon density and $M$ denotes the absolute magnitude of SNe~Ia. Uniform priors are imposed as $\Omega_{\rm m0}\in[0.2,0.5]$, $h\in[0.5,0.8]$, $\omega_{\rm b}\in[0.005,0.05]$ and $M\in[-22,-15]$. The same priors on the common cosmological parameters are adopted for all extended dark energy models. For $+$Union3, the zero point is fixed by \textsc{unity1.5}, so no offset is sampled. For $+$DESY5, the SN data fix the absolute scale only up to an arbitrary additive offset, and we marginalize over this offset analytically (Appendix~\ref{sec:Likelihoods}). We note that $r_{\rm d}$ is not sampled as an independent parameter in our analysis; it is instead computed from $\Omega_{\rm m0}$, $h$, and $\omega_{\rm b}$ via Eq.~\eqref{eq:rd_brieden}, consistent with the standard early-universe relations assumed in constructing the Planck compressed distance priors we adopt (Appendix~\ref{sec:Likelihoods}).

The PL and MPL parametrizations introduce two additional dark energy parameters, $\alpha$ and $w_0$. For the PL model, we adopt uniform priors $\alpha\in[-1, 2]$ and $w_0\in[-2, 1]$ for all data combinations. For the MPL model, we use $\alpha\in[-1,6]$ and $w_0\in[-2,1]$. 

For the PL model, large positive values of $\alpha$ lead to a rapidly evolving equation of state at high redshifts. Although current data provide limited direct constraints in this regime, such behaviour can indirectly influence the expansion history and hence mildly impact the posterior distribution. The prior range of $\alpha$ is chosen such that it does not introduce pathological features in the parameter estimation.

The CPL, BA, and EXP models share the same dark energy parametrization in terms of $w_0$ and $w_{\rm a}$. For these models, we impose uniform priors $w_0\in[-2,0]$ and $w_{\rm a}\in[-2,1]$ for all data sets.

We perform a Markov Chain Monte Carlo (MCMC) analysis to constrain the model parameters using the publicly available {\tt EMCEE} sampler \citep{Foreman-Mackey:2012any}. The resulting chains are analysed and visualised with the {\tt GetDist} package \citep{Lewis:2019xzd}. Convergence of the MCMC chains is assessed using the Gelman--Rubin statistic \citep{Gelman:1992zz}, requiring $|R-1|\lesssim0.01$.

\subsection{Model selection and significance}
\label{sec:tension}

To assess and compare the performance of different dark energy models, we employ standard model selection criteria \citep{Liddle:2006tc,Shi:2012ma,Trotta:2017wnx}, namely the Akaike Information Criterion (AIC) \citep{Akaike74} and the Bayesian Information Criterion (BIC) \citep{Schwarz:1978tpv}. These are defined as ${\rm AIC}=\chi^2_{\rm min}+2N_{\rm para}$ and ${\rm BIC}=\chi^2_{\rm min}+N_{\rm para}\ln N_{\rm data}$, where $\chi^2_{\rm min}$ is the minimum chi-squared value, $N_{\rm para}$ denotes the number of free parameters, and $N_{\rm data}$ is the total number of data points. The relative performance of a model with respect to $\Lambda$CDM is quantified through $\Delta{\rm AIC}={\rm AIC}_{\rm model}-{\rm AIC}_{\Lambda{\rm CDM}}$ and $\Delta{\rm BIC}={\rm BIC}_{\rm model}-{\rm BIC}_{\Lambda{\rm CDM}}$, with negative (positive) values indicating preference for (disfavouring of) the model relative to $\Lambda$CDM.

We quantify the departure of each DDE model from $\Lambda$CDM with a likelihood-ratio test. $\Lambda$CDM is nested in every DDE model considered here. It is recovered at $(w_0,w_{\rm a})=(-1,0)$ for CPL, BA and EXP, and at $(\alpha,w_0)=(0,-1)$ for PL and MPL. Both points lie inside the prior ranges. If $\Lambda$CDM is true, Wilks' theorem \citep{Wilks:1938} states that $\Delta\chi^2_{\rm min}=\chi^2_{\rm min,\Lambda CDM}-\chi^2_{\rm min,DDE}$ follows a $\chi^2$ distribution with $k=2$ degrees of freedom. The $p$-value is $p=\exp(-\Delta\chi^2_{\rm min}/2)$, and we express it as an equivalent Gaussian significance, $\sigma_{\rm LR}=\Phi^{-1}(1-p/2)$, where $\Phi^{-1}$ is the inverse cumulative distribution function of the standard normal distribution. This test does not assume Gaussian posteriors. The $\Lambda$CDM and DDE fits use slightly different compressed Planck priors (Appendix~\ref{sec:Likelihoods}), so $\Delta\chi^2_{\rm min}$ is approximate at that level.

\begin{figure*}[t]
\centering
\includegraphics[scale=0.32]{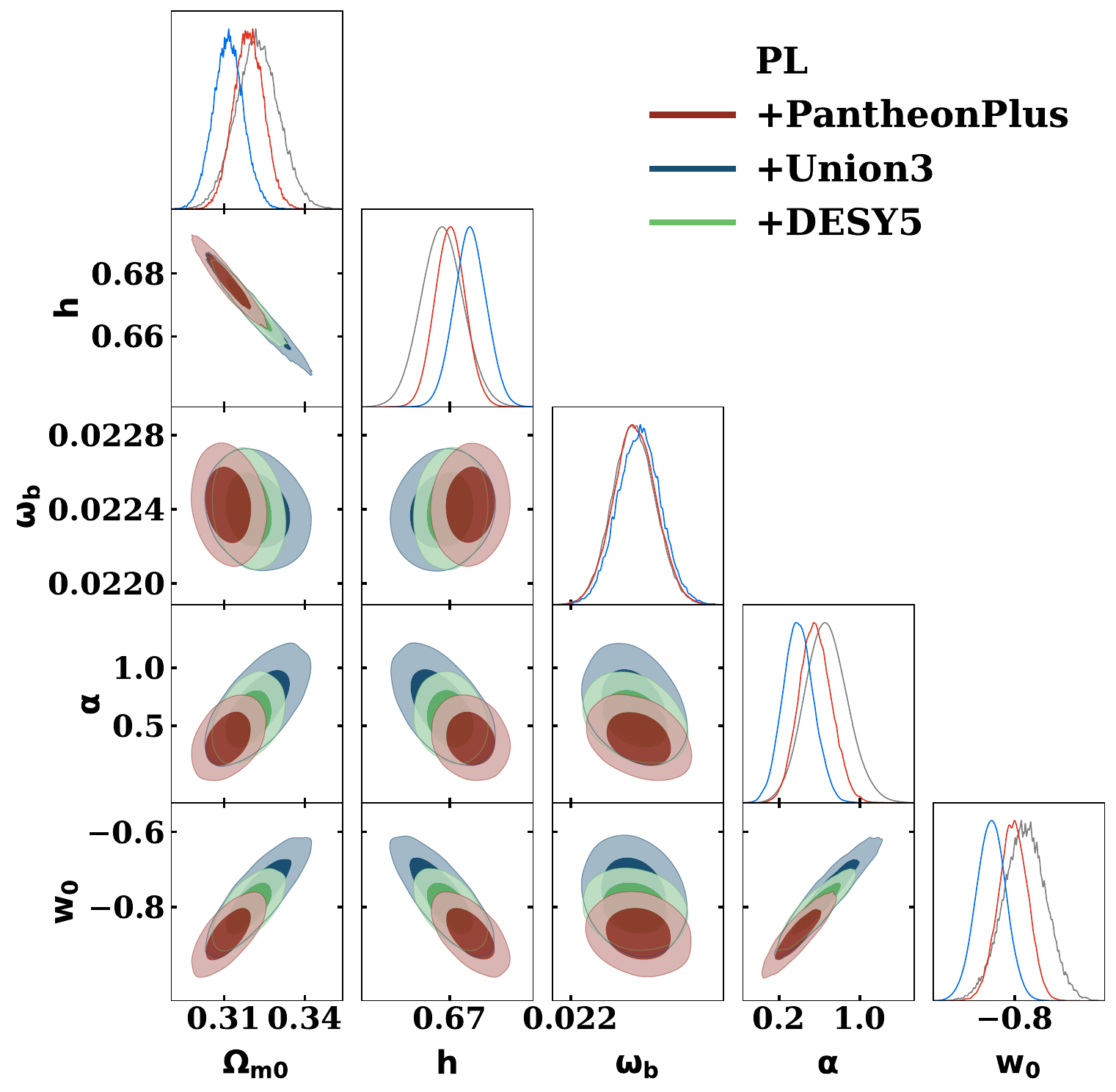}~~~
\includegraphics[scale=0.32]{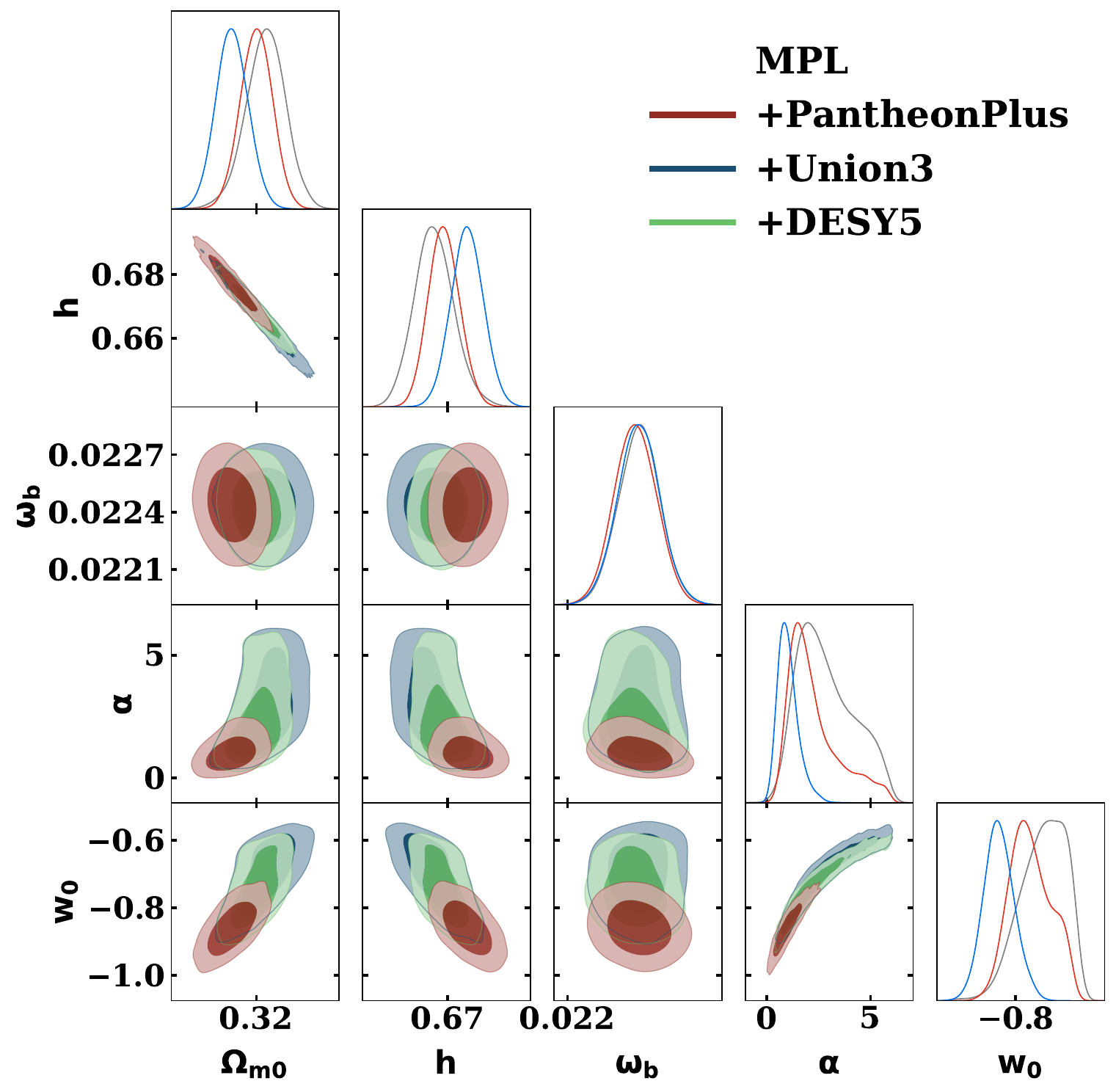}
\caption{$1\sigma$ and $2\sigma$ confidence contours for the parameters of PL (left) and MPL (right) models.
}
\label{fig:constraint}
\end{figure*}

\begin{figure*}[ht]
\centering
\includegraphics[scale=0.45]{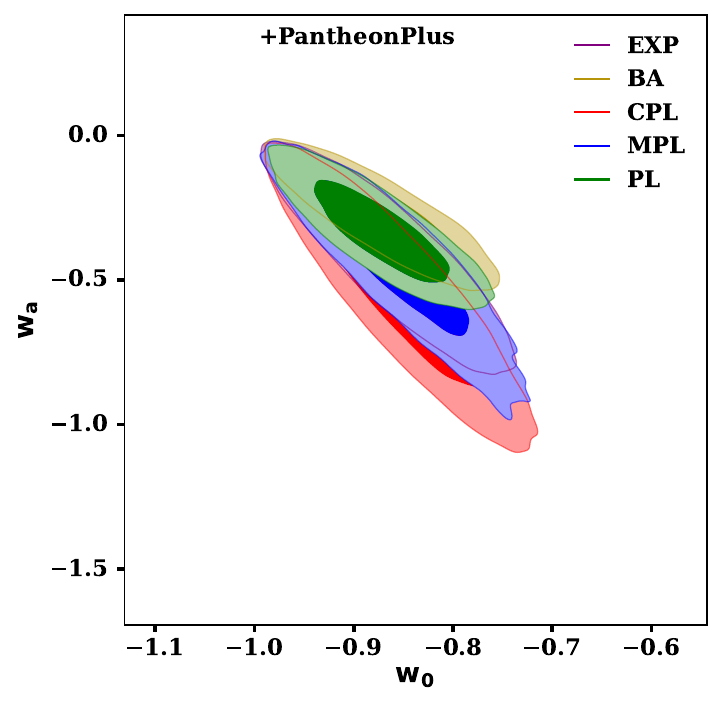}
\includegraphics[scale=0.45]{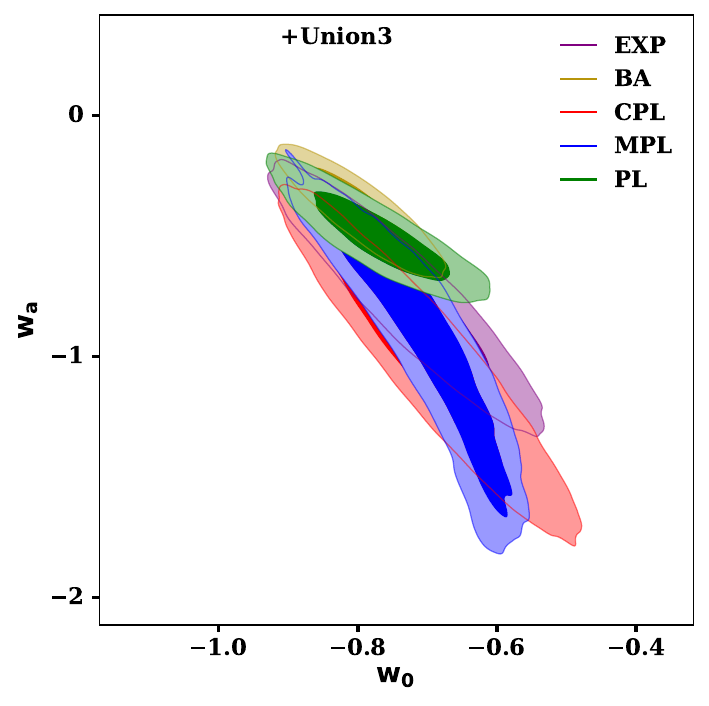}
\includegraphics[scale=0.45]{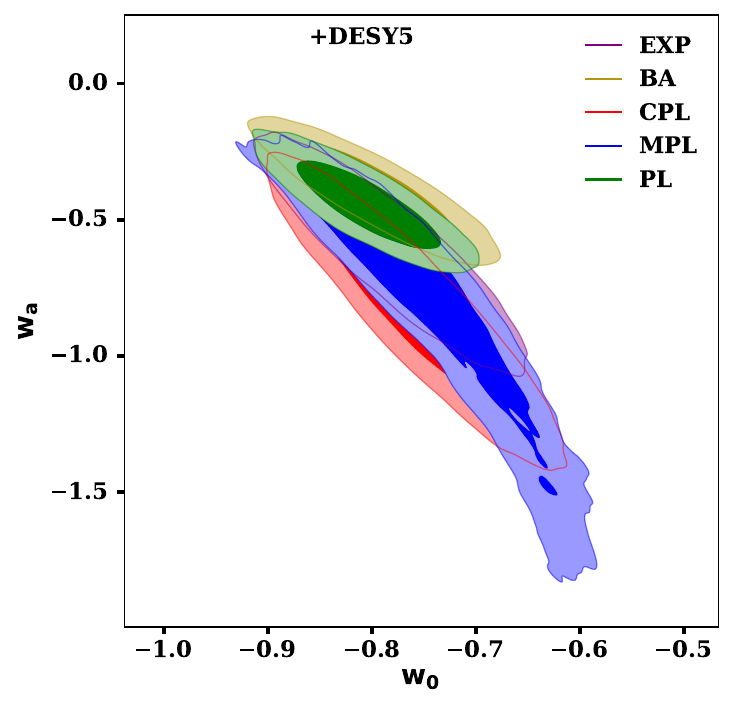}\\
\includegraphics[scale=0.45]{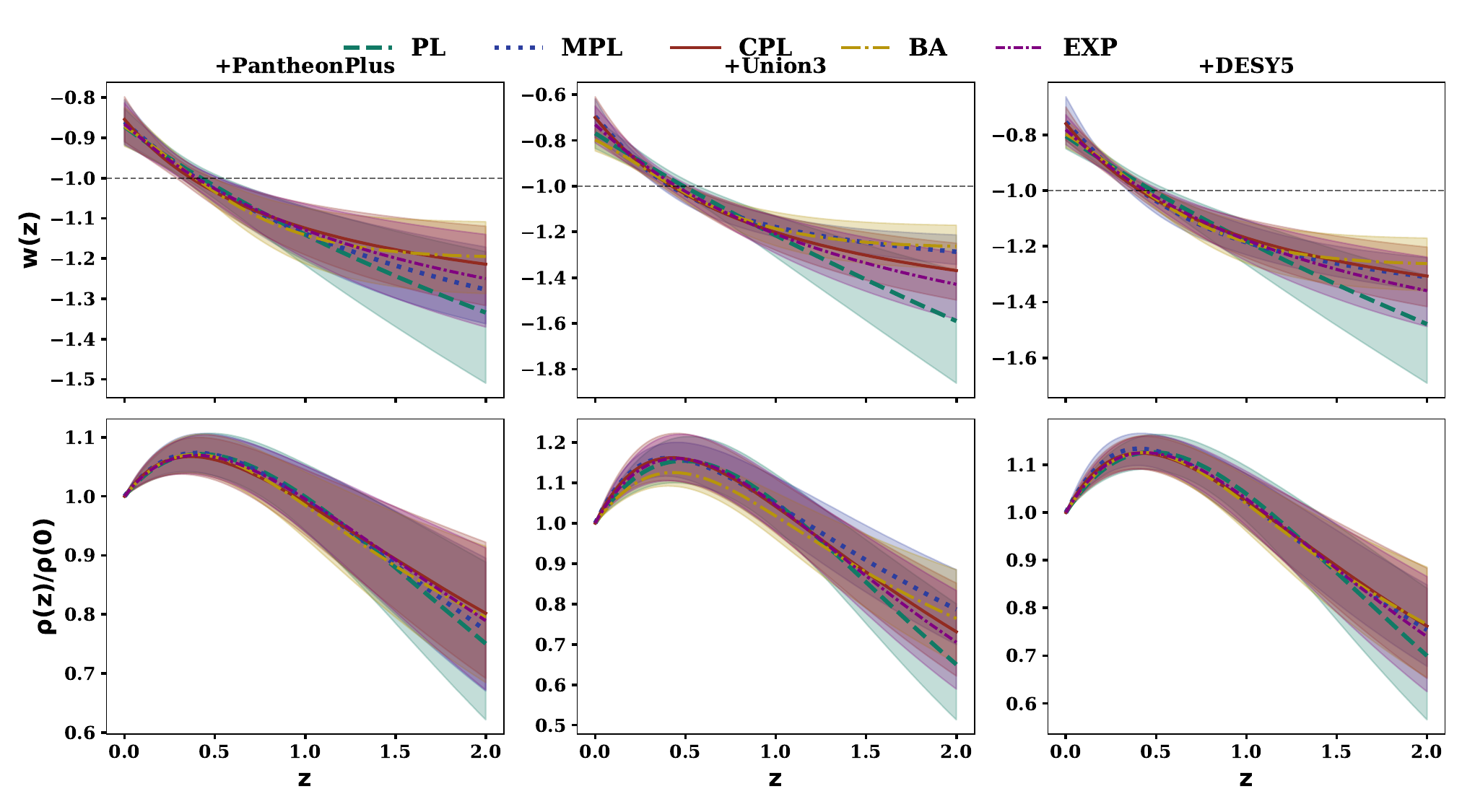}
\caption{Upper: $1\sigma$ and $2\sigma$ confidence contours for the parameters $w_0$ and $w_{\rm a}$ in the PL, MPL, CPL, BA and EXP models. Lower: Reconstructed dark energy EoS $w(z)$ (top row) and normalised dark energy density $\rho(z)/\rho(0)$ (bottom row) for the PL, MPL, CPL, BA, and EXP models, using the $+$PantheonPlus, $+$Union3, and $+$DESY5 data combinations (columns), showing the median evolution (lines) and the corresponding $1\sigma$ credible regions (shaded bands).}
\label{fig:w0-wa}
\end{figure*}

\begin{figure*}[ht]
\centering
\includegraphics[scale=0.41]{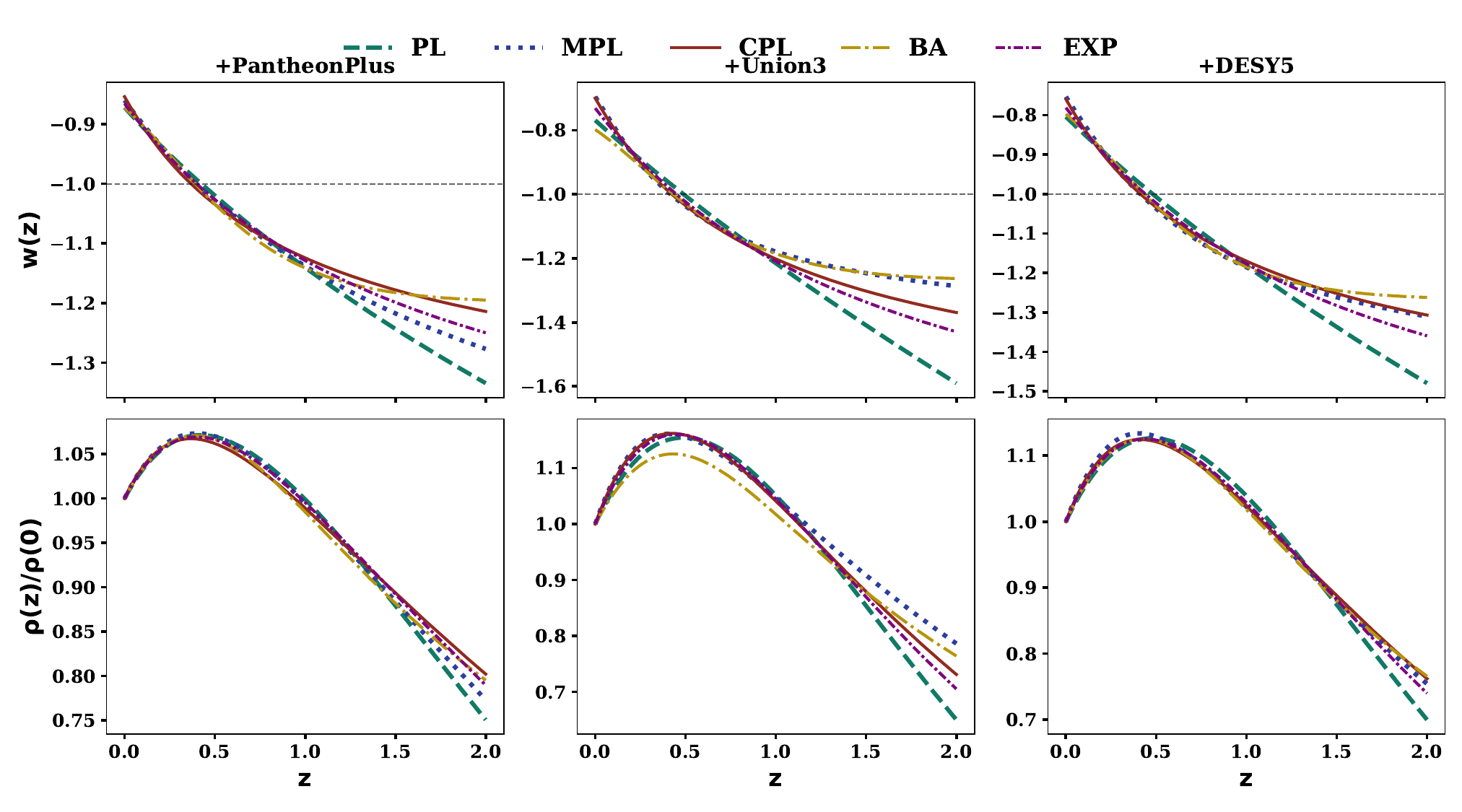}\\ \vskip10pt
\includegraphics[scale=0.45]{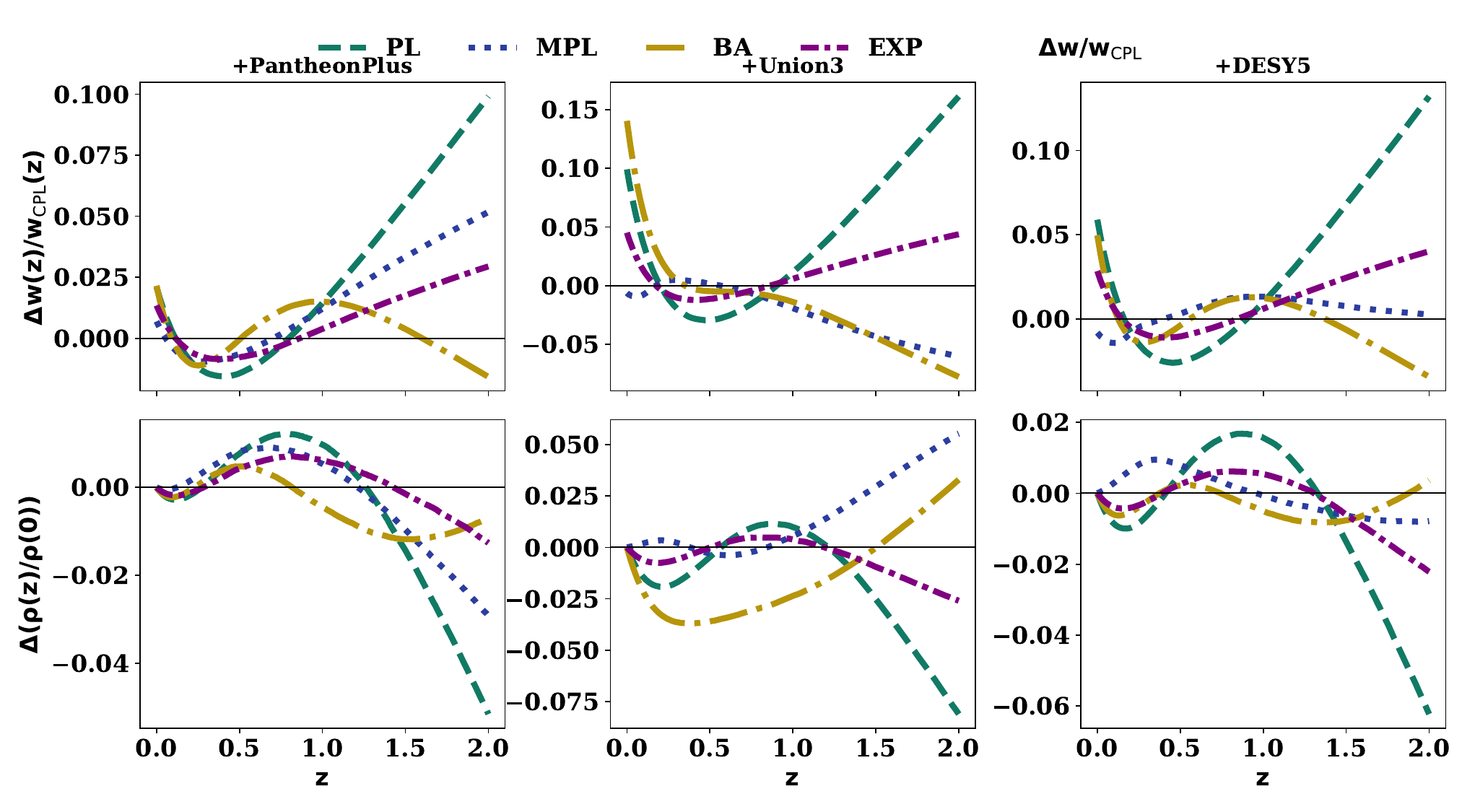}
\caption{Upper: Reconstructed dark energy EoS $w(z)$ (top row) and normalised dark energy density $\rho(z)/\rho(0)$ (bottom row) for the PL, MPL, CPL, BA, and EXP models, using the $+$PantheonPlus, $+$Union3, and $+$DESY5 data combinations (columns), showing the median evolution (lines). Lower: Fractional difference $\Delta w(z)/w_{\rm CPL}(z)$ (top row) and difference $\Delta(\rho(z)/\rho(0))$ (bottom row) of the PL, MPL, BA, and EXP models relative to the CPL evolution, where $\Delta \equiv {\rm CPL} - {\rm Model}$.}
\label{fig:wz_error_bestfit_only}
\end{figure*}

\section{Results}
\label{sec:results}

The Table~\ref{tab:params_constraints} summarises the cosmological parameter constraints for $\Lambda$CDM and the DDE models considered in this work, namely PL, MPL, CPL, BA, and EXP, using the +PantheonPlus, +Union3, and +DESY5 data combinations. Across all datasets, the background parameters $\Omega_{\rm m0}$, $H_0$, $\omega_{\rm b}$ and the absolute magnitude $M$ are broadly consistent among the models, with only minor variations. The contours of the common parameters for the PL and MPL models are shown in Fig.~\ref{fig:constraint}. As illustrated in Fig.~\ref{fig:constraint}, the introduction of DDE primarily affects the dark energy sector, leaving the background parameters largely unchanged. 

The upper panel of Fig.~\ref{fig:w0-wa} shows the $1\sigma$ and $2\sigma$ confidence contours in the $w_0$–$w_{\rm a}$ plane for the different DDE models considered, while the lower panel presents the reconstructed EoS and the normalised energy density, each with their corresponding $1\sigma$ uncertainties, obtained using the best-fit values of the model parameters. For PL and MPL, the $w_{\rm a}$ shown here is a derived quantity obtained from the sampled $(w_0,\alpha)$ via Eqs.~\eqref{eq:wa_PL}--\eqref{eq:wa_MPL}, rather than an independently fitted parameter as in CPL, BA, and EXP. The PL/MPL contours in the $w_0$–$w_{\rm a}$ plane should therefore be read as a projection of a two-parameter posterior, not as a direct constraint on $w_{\rm a}$ itself. Fig.~\ref{fig:wz_error_bestfit_only} shows the evolution of the EoS and the normalised energy density for the different DDE models using their respective best-fit values (upper panel), along with the corresponding differences relative to the CPL evolution (lower panel). A negative difference indicates a stronger or {\it enhanced} phantom behaviour relative to the CPL model. Both figures, Figs.~\ref{fig:w0-wa} and \ref{fig:wz_error_bestfit_only}, show that different parametrizations lead to systematically different best-fit evolutionary trajectories, particularly at intermediate redshifts, although the reconstructed $1\sigma$ credible bands of the different models remain mutually consistent over this redshift range.

Although the CPL and MPL models generally favour more negative best-fit values of $w_{\rm a}$ than the PL model, their evolution is intrinsically restricted by their asymptotic behaviour. In contrast, parametrizations such as PL, which do not impose strong asymptotic constraints, allow a more pronounced departure from $w=-1$ at $z \gtrsim 1$. This leads to a hierarchy in the strength of the best-fit phantom behaviour across models, although the corresponding $1\sigma$ confidence regions in Fig.~\ref{fig:w0-wa} overlap substantially between models.

This hierarchy in the best-fit values is only mildly and not decisively reflected in the statistical inference summarised in Table~\ref{tab:params_stats}. All DDE parametrizations yield improvements in the minimum $\chi^2$ relative to $\Lambda$CDM. The corresponding differences in the AIC indicate a consistent preference for DDE scenarios, while the BIC, as expected, penalises the additional parameters more strongly and reduces the statistical significance of this preference. The likelihood-ratio test gives $\sigma_{\rm LR}=2.2$--$2.5$ for +PantheonPlus, $3.1$--$3.2$ for +Union3 and $3.8$--$3.9$ for +DESY5 (Table~\ref{tab:params_stats}). For a given dataset, the five parametrizations differ by at most $0.25\sigma$. The preference over $\Lambda$CDM therefore depends more on the supernova sample than on the parametrization.

More importantly, within the precision of current data, the various DDE parametrizations yield statistically comparable fits, with only a weak indication that models admitting stronger phantom evolution may provide marginally better fits. For the $+$PantheonPlus data combination, Fig.~\ref{fig:wz_error_bestfit_only} shows that the PL model exhibits the most enhanced phantom behaviour for $z \gtrsim 1$, followed by MPL, EXP, BA, and CPL. The same ordering is reflected in the statistical results, with PL providing the largest improvement in $\chi^2$, followed by MPL, EXP, BA, and CPL. 

This trend persists across the other datasets. For the $+$Union3 combination, the hierarchy from most to least phantom behaviour is PL, EXP, CPL, MPL, and BA, while for $+$DESY5 it is PL, EXP, MPL, CPL, and BA. This ordering broadly tracks the statistical preference summarised in Table~\ref{tab:params_stats}.

Given that current observational constraints on $w(z)$ weaken at higher redshifts, the detailed behaviour of the EoS in this regime is influenced by the chosen parametrization. Nevertheless, a consistent pattern emerges across independent datasets, parametrizations that allow stronger deviations from $w=-1$ at intermediate redshifts tend to yield slightly improved fits to the data. This behaviour is most apparent around $z \sim 1$--2, where the combination of BAO and SNe~Ia provides the dominant sensitivity to dark energy evolution.

However, the magnitude of these improvements remains modest, and the differences between parametrizations are not statistically significant. As a result, the observed hierarchy in phantom behaviour should not be interpreted as evidence for a preferred model or for a definitive detection of enhanced phantom dynamics. Rather, it indicates that current data are compatible with a range of dynamical dark energy behaviours, and that parametrizations with greater flexibility can accommodate features that are not strongly constrained by existing observations.

\section{Conclusions}
\label{sec:conc}

In this work, we have investigated to what extent current cosmological observations allow or favour phantom-like behaviour in dynamical dark energy models, and how such inferences depend on the chosen parametrization. Using the $+$PantheonPlus, $+$Union3, and $+$DESY5 data combinations, we performed a comprehensive comparison between $\Lambda$CDM and five DDE parametrizations, PL, MPL, CPL, BA, and EXP. Our results show that all DDE models yield modest improvements in the fit relative to $\Lambda$CDM (Table~\ref{tab:params_stats}), leading to reductions in $\chi^2_{\rm min}$ and mildly negative $\Delta$AIC values. The BIC penalises the additional parameters more strongly and does not provide decisive evidence for departures from $\Lambda$CDM.

A qualitative hierarchy emerges when the reconstructed equations of state are examined (Fig.~\ref{fig:wz_error_bestfit_only}). The PL parametrization systematically exhibits the most enhanced phantom behaviour for $z \gtrsim 1$, followed, depending on the dataset, by MPL or EXP, while CPL and BA typically display comparatively milder departures. This ordering is broadly reflected in the statistical results. However, we find that it is largely driven by the intrinsic extrapolation properties of the parametrizations rather than by strong constraints from the data. While the statistical preference remains modest, the persistence of this hierarchy across all three data combinations indicates a coherent trend in how different parametrizations capture the allowed dark energy evolution, though the reconstructed EoS and density evolutions of the different models remain mutually consistent within $1\sigma$. The likelihood-ratio test gives a significance of $2.2$--$2.5\sigma$ for +PantheonPlus, $3.1$--$3.2\sigma$ for +Union3 and $3.8$--$3.9\sigma$ for +DESY5. For a given dataset the value is nearly the same for all five parametrizations, so the preference depends more on the supernova sample than on the parametrization.

Overall, our findings suggest that present observations primarily constrain the low-redshift behaviour of dark energy, while the evolution at higher redshifts remains weakly constrained and depends on the assumed parametrization. Models with restricted asymptotic behaviour, such as CPL and BA, limit the extent of phantom evolution, whereas parametrizations such as PL, which are less constrained at high redshift, allow more pronounced departures from $w=-1$ at $z\gtrsim1$. These differences change the fit only slightly. The minimum $\chi^2$ of PL and CPL, which have the same number of free parameters, differs by at most 1.3 for all three data combinations.

Relative to $\Lambda$CDM, all five parametrizations show a modest preference for dynamical dark energy with phantom crossing in the recent past, consistent with recent DESI analyses. The AIC favours the DDE models for all three SNe~Ia samples. The BIC mildly favours them for $+$Union3 and $+$DESY5 ($\Delta\mathrm{BIC}=-4.7$ to $-2.5$) and disfavours them for $+$PantheonPlus ($\Delta\mathrm{BIC}=+6.3$ to $+7.6$). The preference therefore depends more on the SNe~Ia sample than on the parametrization, and, since it appears in all five forms, it is not tied to a specific functional form. The strength of the phantom evolution at intermediate redshifts is not robustly constrained by current data. Future observations with higher precision are needed to determine whether these features reflect genuine dark energy dynamics or arise from parametrization-dependent effects.

\appendix
\section{Likelihoods}
\label{sec:Likelihoods}

\subsection{Compressed CMB likelihood}
\label{sec:cmb_likelihood}

We employ the compressed Planck 2018 distance priors~\citep{2019JCAP}, which encapsulate the main geometric
information from the CMB anisotropy measurements through the parameter
vector
\begin{equation}
\mathbf{p} = \left( R,\, \ell_A,\, \omega_b \right)^{\rm T},
\label{eq:cmb_vector}
\end{equation}
where the shift parameter $R$ and acoustic scale $\ell_A$ are defined as
\begin{eqnarray}
R &=&\sqrt{\Omega_{m0}}\,\frac{H_0D_M(z_*)}{c}\, , \\
\ell_A &=& \pi\,\frac{D_M(z_*)}{r_s(z_*)}\, ,
\end{eqnarray}
with
\begin{eqnarray}
D_M(z) &=& \int_0^z \frac{c\,dz'}{H(z')}\, ,
\label{eq:DM}\\
r_s(z) &=& \int_z^\infty
\frac{c_s(z')}{H(z')}dz' \, ,
\end{eqnarray}
where $D_M(z)$ is the transverse comoving distance to the redshift $z$, $r_s(z)$ is the sound horizon, $z_*$ is the photon-decoupling epoch and $c_s(z)$ is the sound speed in the photon--baryon fluid.

The likelihood is assumed to be Gaussian,
\begin{equation}
\chi^2_{\rm CMB} = \Delta^{\rm T} \mathbf{C}^{-1} \Delta ,
\label{eq:chi2_cmb}
\end{equation}
where
\begin{equation}
\Delta = \mathbf{p}_{\rm th} - \mathbf{p}_{\rm obs},
\end{equation}
where $\mathbf{p}_{\rm th}=(R,\ell_A,\omega_b)^{\rm T}$ denotes the vector of theoretically predicted compressed CMB observables for a given cosmological model, while $\mathbf{p}_{\rm obs}$ is the corresponding vector of Planck 2018 distance priors and $\mathbf{C}$ is the covariance matrix reconstructed from the
correlation coefficients and uncertainties provided in
Ref.~\citep{2019JCAP}.

Following Ref.~\citep{2019JCAP},
we employ the $\Lambda$CDM compressed priors when analysing
$\Lambda$CDM and the $w$CDM compressed priors for all
non-$\Lambda$CDM dark-energy parametrizations. For the $\Lambda$CDM model
\begin{equation}
\mathbf p_{\rm obs}^{\Lambda{\rm CDM}}
=
\begin{pmatrix}
1.7502\\
301.471\\
0.02236
\end{pmatrix}.
\end{equation}
with correlation matrix
\begin{equation}
\mathbf R=
\begin{pmatrix}
1.00 & 0.46 & -0.66\\
0.46 & 1.00 & -0.33\\
-0.66 & -0.33 & 1.00
\end{pmatrix},
\end{equation}
and standard deviations
\begin{equation}
\boldsymbol{\sigma} = (0.0046,\, 0.089,\, 0.00015).
\end{equation}

For dynamical dark-energy models we instead employ the $w$CDM
compressed distance priors of Ref.~\citep{2019JCAP},
\begin{equation}
\mathbf p_{\rm obs}^{w{\rm CDM}}
=
\begin{pmatrix}
1.7493\\
301.462\\
0.02239
\end{pmatrix}.
\end{equation}
with correlation matrix
\begin{equation}
\mathbf R=
\begin{pmatrix}
1.00 & 0.47 & -0.66\\
0.47 & 1.00 & -0.34\\
-0.66 & -0.34 & 1.00
\end{pmatrix}.
\end{equation}
and standard deviations
\begin{equation}
\boldsymbol{\sigma}
=
(0.0046,\,
0.089,\,
0.00015).
\end{equation}
We use the $w$CDM compressed priors for all DDE models. These priors were derived for a constant equation of state, and we have not tested them against the full Planck likelihood for the DDE parametrizations considered here. This is a systematic limitation of our analysis.

\subsection{DESI DR2 BAO likelihood}
\label{sec:DESI_likelihoods}
We use the compressed DESI DR2 BAO likelihood released by the DESI collaboration~\citep{DESI:2025zgx}, consisting of thirteen effectively independent BAO observables. The corresponding mean vector and covariance matrix are publicly available through the Cobaya BAO repository\footnote{\url{https://github.com/CobayaSampler/bao_data/tree/master/desi_bao_dr2}}. Assuming a Gaussian likelihood, the corresponding $\chi^2$ statistic is
\begin{equation}
\chi^2_{\rm DESI\,DR2} =
\Delta^{\rm T}\mathbf{C}^{-1}\Delta,
\label{eq:chi2_desi_dr2}
\end{equation}
where $\mathbf{C}$ is the published covariance matrix and
\begin{equation}
\Delta = \mathbf{d}_{\rm th} - \mathbf{d}_{\rm obs},
\end{equation}
with $\mathbf{d}_{\rm th}$ denoting the vector of theoretically predicted BAO observables for a given cosmological model, while $\mathbf{d}_{\rm obs}$ is the corresponding vector of DESI DR2 measurements, given by~\citep{DESI:2025zgx}
\begin{equation}
\mathbf{d}_{\rm obs}
=
\left(
\begin{array}{c}
D_V(0.295)/r_d \\
D_M(0.51)/r_d \\
D_H(0.51)/r_d \\
D_M(0.706)/r_d \\
D_H(0.706)/r_d \\
D_M(0.934)/r_d \\
D_H(0.934)/r_d \\
D_M(1.321)/r_d \\
D_H(1.321)/r_d \\
D_M(1.484)/r_d \\
D_H(1.484)/r_d \\
D_M(2.33)/r_d \\
D_H(2.33)/r_d
\end{array}
\right)
=
\left(
\begin{array}{c}
7.94167639\\
13.58758434\\
21.86294686\\
17.35069094\\
19.45534918\\
21.57563956\\
17.64149464\\
27.60085612\\
14.17602155\\
30.51190063\\
12.81699964\\
38.98897396\\
8.63154567
\end{array}
\right),
\end{equation}

For the spatially flat cosmologies considered in this work, apart from the transverse comoving distance $D_M(z)$, defined in Eq.~\eqref{eq:DM}, the BAO distance measures entering the likelihood are
\begin{eqnarray}
D_H(z) &=& \frac{c}{H(z)}, \\
D_V(z) &=& \left[z\,D_H(z)\,D_M^2(z)\right]^{1/3},
\end{eqnarray}
where $r_d$ denotes the sound horizon at the drag epoch. Assuming a standard pre-recombination history, the sound horizon is well approximated by the fitting formula of Ref.~\citep{Brieden:2022heh} in terms of the baryon and CDM densities, photons, and other relativistic components,
\begin{align}
    r_{\rm d} =&  147.05~{\rm Mpc} \left(\frac{\Om_{\m0}h^2}{0.1432}\right)^{-0.23} \left(\frac{\Om_{\rm b 0}h^2}{0.02236}\right)^{-0.13} \nn \\ & \times \left(\frac{N_{\rm eff}}{3.04}\right)^{-0.1} \, .
    \label{eq:rd_brieden}
\end{align}
For our analysis we consider $N_{\rm eff}=3.04$ for simplicity.

\subsection{Type Ia supernova likelihoods}
\label{sec:sn_likelihoods}

We use three independent Type Ia supernova compilations, PantheonPlus~\citep{Scolnic:2021amr,Brout:2022vxf}, Union3~\citep{Rubin:2023ovl}, and DESY5 sample~\citep{DES:2024jxu} with its latest DES-Dovekie recalibration \cite{DES:2025sig}. Each compilation releases a different light curve observable. PantheonPlus and Union3 release apparent magnitudes $m_{B,i}$, while DESY5 releases distance moduli $\mu_{{\rm obs},i}$. The distance modulus is defined as
\begin{equation}
\mu_{\rm th}(z_i) = 5\log_{10}\!\left[\frac{D_L(z_i)}{\rm Mpc}\right] + 25 ,
\label{eq:mu_th_general}
\end{equation}
where $D_L(z)$ is the model luminosity distance. For a compilation with $N$ data points at redshifts $z_i$, we denote by $\mathbf{C}$ the full statistical plus systematic covariance matrix released with it. All three $\chi^2$ statistics below are evaluated via the Cholesky factorization $\mathbf{C}=\mathbf{L}\mathbf{L}^{\top}$, where $\mathbf{L}$ is the unique lower-triangular matrix with strictly positive diagonal entries satisfying this decomposition,\footnote{Existence and uniqueness of $\mathbf{L}$ require $\mathbf{C}$ to be symmetric positive-definite.} followed by triangular forward/backward substitution rather than explicit inversion of $\mathbf{C}$, for numerical stability and reduced computational cost. The three compilations differ in the observable used and in how the SN Ia absolute magnitude/$H_0$ degeneracy is handled.

\paragraph{PantheonPlus (PP).} We apply the cut $z>0.01$ to the full sample of 1701 light curves, to remove SNe dominated by peculiar velocity noise. The Cepheid-host (calibrator) SNe are also removed from the data vector and the covariance matrix. This leaves $1590$ light curves, with the data vector and the released statistical plus systematic covariance matrix cut consistently. The released observable $m_{B,i}$ is the bias corrected apparent magnitude. So, PantheonPlus does not fix the SNe~Ia absolute magnitude/$H_0$ degeneracy. We constrain the parameter $M$ along with the cosmological parameters and compare $m_{B,i}$ with the theoretically calculated apparent magnitude $m_{B,\rm th}(z_i) = \mu_{\rm th}(z_i) + M$,
\begin{align}
\chi^2_{\rm PP} &= \boldsymbol{\Delta}^{\!\top}\mathbf{C}^{-1}\boldsymbol{\Delta} \\ 
\Delta_i &= m_{B,i} - m_{B,\rm th}(z_i) = m_{B,i} - \mu_{\rm th}(z_i) - M .
\label{eq:chi2_pantheon}
\end{align}

\paragraph{Union3.} The Union3 sample consists of 2087 SNe Ia from 24 surveys, compressed via the \textsc{unity1.5} Bayesian hierarchical framework into $N=22$ redshift-binned apparent magnitude estimates $m_{B,i}$ over $0.05\lesssim z\lesssim2.26$. Unlike the PantheonPlus $m_{B,i}$, the Union3 $m_{B,i}$ has its absolute magnitude zero point fixed internally by \textsc{unity1.5}, so it is already on the same additive scale as $\mu_{\rm th}(z_i)$. No nuisance parameter is sampled, and
\begin{equation}
\chi^2_{\rm Union3} = \boldsymbol{\Delta}^{\!\top}\mathbf{C}^{-1}\boldsymbol{\Delta}, \qquad \Delta_i = m_{B,i} - \mu_{\rm th}(z_i) .
\label{eq:chi2_union3}
\end{equation}

\paragraph{DES-Dovekie (DESY5).} The DES-Dovekie recalibrated DESY5 sample comprises approximately 1600 DES SNe and about 200 low redshift SNe from external surveys, for which the released observable is the distance modulus $\mu_{{\rm obs},i}$ with per SN uncertainty $\sigma_{\mu,i}$. The covariance matrix is $\mathbf{C}=\mathbf{C}_{\rm sys}+\mathrm{diag}(\sigma_{\mu,i}^2)$, where $\mathbf{C}_{\rm sys}$ is the released systematic covariance and the diagonal term adds the statistical uncertainty of each SN.

Two redshifts enter the luminosity distance. $z_{\rm hel}$ is the heliocentric redshift, measured directly from the observed spectrum. $z_{\rm cmb}$ is the redshift in the CMB rest frame, corrected for the Sun's motion and, for nearby SNe, for peculiar velocities. The luminosity distance is
\begin{equation}
D_L(z_{\rm cmb},z_{\rm hel}) = (1+z_{\rm cmb})(1+z_{\rm hel})\,D_A(z_{\rm cmb}) .
\end{equation}
The $(1+z_{\rm hel})$ factor accounts for photon time dilation as seen from Earth. The $(1+z_{\rm cmb})$ factor and $D_A(z_{\rm cmb})$ come from the cosmological distance redshift relation, since $z_{\rm cmb}$ is the redshift that traces the Hubble flow. Rather than sampling $M$ explicitly, we marginalize over the additive offset analytically, assuming a flat prior \citep{Goliath:2001af,Lazkoz:2005sp}, with $\Delta_i = \mu_{{\rm obs},i} - \mu_{\rm th}(z_i)$,
\begin{eqnarray}
\chi^2_{\rm DESY5} &=& \chi_0 - \frac{B^2}{E} + \ln\!\left(\frac{E}{2\pi}\right), \qquad 
\chi_0 = \boldsymbol{\Delta}^{\!\top}\mathbf{C}^{-1}\boldsymbol{\Delta}, \quad \nn \\
B &=& \mathbf{1}^{\!\top}\mathbf{C}^{-1}\boldsymbol{\Delta}, \quad
E = \mathbf{1}^{\!\top}\mathbf{C}^{-1}\mathbf{1} .
\label{eq:chi2_desy5}
\end{eqnarray}
The term $\ln(E/2\pi)$ is the same for all models and cancels in $\Delta\chi^2$.

\subsection{$H(z)$ likelihood}
\label{sec:Hz_likelihood}

We use the compilation of CC  measurements of $H(z)$ at redshifts $z_i$ with observed values $H_{{\rm obs},i}$ and associated uncertainties $\sigma_{H,i}$~\cite{2018JCAP...04..051G}. Assuming independent Gaussian errors, the likelihood is

\begin{equation}
\chi^2_{H(z)}
=
\sum_{i=1}^{N}
\frac{\left[
H_{{\rm obs},i}
-
H_{\rm th}(z_i)
\right]^2}
{\sigma_{H,i}^{\,2}},
\label{eq:chi2_Hz}
\end{equation}

where $H_{\rm th}(z)$ denotes the theoretical Hubble parameter predicted by the cosmological model under consideration.

\bibliographystyle{apsrev4-2} 
\bibliography{references}

\end{document}